\documentclass[]{spie}  

\usepackage{amsmath,amsfonts,amssymb}
\usepackage{graphicx}
\usepackage[colorlinks=true, allcolors=blue]{hyperref}
\usepackage{tikz}
\usetikzlibrary{arrows.meta,positioning,fit,shapes.geometric}
\usepackage{algorithm}
\usepackage{indentfirst}
\usepackage{algpseudocode}

\title{Out-of-Band Power Side-Channel Detection for Semiconductor Supply Chain Integrity at Scale}

\author{Rajiv Thummala}
\author{Katherine Winton}
\author{Luke Flores}
\author{Elizabeth Redmond}
\author{Gregory Falco}

\affil{Cornell University}
\affil{Brooks Tech Policy Institute}
\affil{Ithaca, NY, USA}

\authorinfo{This work was supported by the U.S. Department of Defense (DoD) under Award No. HQ00342520002 as
part of the Semiconductor Supply Chain and Cybersecurity Assessment program (CFDA 12.599).
}

\begin{document} 
\maketitle

\begin{abstract}
Out-of-band screening of microcontrollers is a major gap in semiconductor supply chain security. High-assurance techniques such as X-ray and destructive reverse engineering are accurate but slow and expensive, hindering comprehensive detection for hardware Trojans or firmware tampering. Consequently, there has been increased interest in applying machine learning techniques to automate forensic examination, enabling rapid, large-scale inspection of components without manual oversight. We introduce a non-destructive screening method that uses power side-channel measurements and generative modeling to detect tampering in commodity microcontrollers without trusted hardware. As a proof-of-concept, differential power analysis (DPA) traces are collected from the ChipWhisperer and a generative adversarial network (GAN) is trained only on benign measurements to learn nominal power behavior. The trained discriminator then serves as a one-class anomaly detector. We report detection performance on multiple tampering scenarios and discuss how this technique can serve as an intermediate screening tier between basic functional tests and high-cost forensic analysis. The proposed method is evaluated in the context of semiconductor supply chain practice and policy to assess its suitability as an intermediate assurance mechanism.
\end{abstract}



\section{INTRODUCTION}
\label{sec:intro}
Microcontrollers and integrated circuits form the foundation of modern critical systems, yet they are less likely to be subject to rigorous provenance verification or technical validation for hardware integrity before system integration \cite{nist1800-34b2022}. Hardware Trojans inserted during design or manufacturing and counterfeit or remarked components pose significant threats to defense systems, industrial control infrastructure, and safety-critical platforms \cite{rekhi2025analyzing}. While existing supply chain risk management frameworks acknowledge these risks, they provide limited technical controls for rapid, scalable validation of component integrity prior to integration.

Current approaches to hardware assurance present a fundamental tradeoff between thoroughness and scalability. High-assurance techniques such as X-ray computed tomography, decapsulation and delayering, and destructive reverse engineering can detect structural anomalies and unauthorized circuitry with high confidence \cite{Lindgren2021Autoencoder}. However, these forensic methods require days to weeks per device, specialized facilities, and expert analysis, making them economically infeasible for comprehensive screening. Consequently, they are applied only to small statistical samples, leaving the vast majority of components unvalidated \cite{nist1800-34b2022}. Conversely, standard acceptance testing relies on functional verification and boundary scan, which execute quickly but offer negligible protection against integrity modifications that preserves nominal input-output behavior. This leaves systems integrators with no practical method to screen incoming semiconductor components for integrity violations at the scale and speed required by modern supply chains. 

This paper presents a nondestructive screening method that enables rapid, scalable detection of tampering in commodity microcontrollers before system integration. Our approach leverages power side-channel measurements and generative modeling to identify anomalous behavior without requiring trusted hardware features or device modification. We collect differential power analysis (DPA) traces from a microcontroller executing a standardized workload using the ChipWhisperer-lite \cite{ChipWhispererLiteDocs}. A generative adversarial network (GAN) is trained exclusively on power traces from nominal devices to learn the characteristic power consumption profile of benign operation. The trained discriminator then functions as a one-class anomaly detector during screening, flagging devices whose power behavior deviates from the learned baseline as candidates for further investigation. Our proposed method is evaluated within the context of semiconductor supply chain practice and policy to assess deployment feasibility and integration.


\section{SUPPLY-CHAIN ASSURANCE CONTEXT}
\label{sec:supplychain}
%
\subsection{Microelectronics Supply Chain Risk}
Microcontrollers are designed, manufactured, packaged, tested, and shipped through distribution networks that involve many companies and handoffs. Even for routine commercial devices, this path typically crosses multiple firms before a part is integrated into a finished platform. Additional handoffs occur when devices are replaced, reprogrammed, refurbished, or repaired. Prior to integration, it is difficult to produce strong, unit-level evidence that the device is authentic, unmodified, and consistent with what the bill of materials assumes.

Risk begins in the design stage, where third-party building blocks and design artifacts can be incorporated before a chip is fabricated. It continues through fabrication and packaging, where the device is produced and prepared for distribution. It extends into board assembly and platform integration, where parts are programmed, configured, and placed into systems. Finally, it persists in depot maintenance and field sustainment, where replacement parts and firmware updates are common and where provenance can be harder to establish than in initial production.

Hardware Trojans and other unauthorized hardware changes are most plausibly introduced during design or manufacturing, and research has shown that such changes can be engineered to remain quiet during ordinary testing and to activate only under rare conditions\cite{ieee5406669}. Firmware and configuration changes, by contrast, are most plausibly introduced during programming, repair, refurbishment, or depot processes, where legitimate reprogramming activity can obscure unauthorized modification. Counterfeit and remarked components add a related risk wherein a device may be electrically plausible yet not what it claims to be, with reliability and security properties that diverge from expectations.

Responsibility for assurance is distributed. Component suppliers, board manufacturers, and integrators each control different portions of the lifecycle, and no single actor has full visibility. In defense and other high-consequence settings, contractors and system integrators typically carry the burden of demonstrating reasonable assurance to government customers, while government frameworks increasingly expect that risk controls will be documented, repeatable, and tied to system criticality\cite{dfars2522467007}.

\subsection{Existing Assurance Techniques And Scalability Limits}
High-assurance methods can produce strong evidence of authenticity and integrity, but they are difficult to apply broadly. Imaging methods such as X-ray and CT can reveal certain physical irregularities and packaging anomalies. More conclusive approaches include removing the package and inspecting the die and layers, typically destructive, as well as destructive reverse engineering that can confirm structural conformance. Where available, detailed scan-based or specialized functional testing can provide deeper insight into device behavior, and some laboratories apply side-channel analysis as an investigative tool when a part is already suspected.

These approaches can provide clear technical justification for why a part is trusted or rejected. The limiting factors are operational. They require specialized facilities, expensive equipment, skilled analysts, and substantial time per device. Some methods permanently consume the part. As a result, these techniques are usually applied to small statistical samples, to high-risk lots, or to devices already flagged by other indicators rather than to every unit that enters a platform inventory.

Most organizations therefore rely on faster acceptance practices, typically centered on functional checks and standard test interfaces. These are practical for throughput, but they provide limited protection against integrity changes that preserve normal input–output behavior. This is a known failure mode in the hardware Trojan literature: malicious logic can be designed to avoid activation under routine tests and to trigger only under rare internal states\cite{ieee5406669}. In the same way, firmware modifications can preserve outward behavior while changing control flow, privilege, or update pathways that are unlikely to be exercised during typical screening.

The issue is not whether strong techniques exist; it is that the strongest techniques cannot be used at the scale and speed of modern supply chains. This creates a practical assurance discontinuity. Comprehensive screening is infeasible, yet the consequences of a small number of compromised or counterfeit parts can be large once devices are integrated into critical systems.

\subsection{Requirements For A Scalable Screening Tier}
A scalable screening tier is intended to narrow that discontinuity. It is not meant to replace destructive forensics, and it is not satisfied by basic functional testing alone. Instead, it provides a repeatable, non-destructive check that can be applied widely and that produces a decision signal suitable for operational use. Devices can be cleared, held for quarantine, or selected for deeper forensic analysis based on measured deviation rather than broad suspicion.

To fit procurement, incoming inspection, and depot workflows, such screening must work on unmodified commercial devices and must not require special trusted features built into the part. It must be fast enough to support routine handling, with time-per-unit compatible with manufacturing and maintenance environments. It must also produce an auditable output that can be documented and tied to a risk decision, consistent with how supply chain risk management is treated in current guidance and in defense acquisition expectations.

Side-channel screening is attractive under these constraints because it observes the device from the outside while reflecting internal activity. Power measurements, in particular, can provide sensitivity to changes in instruction mix, code paths, and hardware behavior even when basic functionality is preserved. Differential power analysis established that power traces correlate with internal computation and can reveal information not visible through ordinary input–output testing\cite{usenixsecurity23kogler}. More recent work has treated power measurements as an out-of-band signal for detecting abnormal behavior, supporting the broader idea that power can provide integrity-relevant evidence under constrained trust assumptions\cite{cathis2024sok_power_malware}.

This creates an intermediate evidentiary layer: stronger than a simple functional check, but far cheaper and faster than destructive analysis. In the next sections, we evaluate whether modeling nominal power behavior with generative methods can meet these requirements and serve as a practical screening tier for supply-chain assurance.

\section{TECHNICAL BACKGROUND AND RELATED WORK}
\label{sec:background}

\subsection{Hardware Trojan And Counterfeit Detection}
The security and authenticity of integrated circuits (IC) are threatened by two related but distinct problems: hardware Trojans (malicious, stealthy modifications of an IC’s logic or routing) and counterfeit parts (devices, sometimes recycled, whose provenance or integrity has been compromised). Over the past decade, literature has converged around 4 broad categories of detection and mitigation techniques: imaging and physical inspection, layout-level and design-time techniques, on-chip sensors and built-in self-test, and statistical screening via parametric electrical testing. Each category addresses different threat models and has tradeoffs in cost, invasiveness, required access level, and deployment point in the supply chain.

Physical inspection uses direct imaging modalities, such as optical microscopy, X-ray/CT scanning, scanning electron microscopy (SEM), and focused ion beam (FIB) cross-sectioning, often combined with automated image processing and layout overlays to reveal added or modified wiring, extra cells, or anomalous structures in the die or package. These methods provide high-confidence forensic evidence and can expose alterations introduced at a foundry or during rework. However, they are time and resource-costly and frequently destructive or semi-destructive (especially FIB/SEM workflows). They also commonly require golden reference images or layout files for reliable differencing; absent a trusted baseline, interpretation becomes subjective and error-prone. Thus, imaging is most appropriate for targeted forensic analysis or when a small set of suspect devices must be authenticated \cite{trojan_physical_inspection, trojan_self_testing}.

A large body of work focuses on preventing or detecting Trojans during design verification and gate-level analysis. Techniques include controllability and observability metrics, unused-circuit identification, formal information flow checks, Boolean functional analysis, and newer graph-based learning that models netlists as graphs to detect anomalous structure. These approaches can be effective at catching inserted logic before fabrication and can be automated to scale to large netlists. Their central limitation is that they require access to design artifacts (register-transfer level, gate-level netlists, or GDSII) and typically assume the ability to compare against a golden or trusted model; therefore, they do not apply to legacy devices or closed-source commercial off-the-shelf (COTS) components acquired downstream from the original equipment manufacturer \cite{trojan_prevention, trojan_neural_networks}.

Embedding sensors (e.g., ring-oscillator arrays, current/voltage monitors, temperature sensors) and expanding built-in self-test (BIST) functionality can enable in-field or manufacturing-time detection of malicious activity or anomalous operational signatures. Because these schemes can monitor runtime behavior and trigger conditions, they are attractive for high-assurance systems. The drawback is that they require design cooperation and firmware support, so they are not feasible for off-the-shelf or already-deployed parts. Moreover, adaptive adversaries may design Trojans that avoid triggering these monitors or that manipulate sensor outputs, and benign effects such as aging or environmental variation complicate signal interpretation \cite{trojan_self_testing}.

Counterfeit detection often relies on statistical analysis of electrical parameters measured in manufacturing or lab tests: IDDQ/leakage, timing/frequency distributions, threshold voltages, and analog parametrics obtained from controlled sweeps. By comparing device measurements to expected population distributions or golden baselines, outliers corresponding to recycled, remarked, or counterfeit parts can be flagged. These methods offer higher throughput than deep imaging and can be non-destructive, making them attractive for supply-chain screening. Their limitations stem from process and device variability, aging, and overlap between the signatures of genuine but aged parts and counterfeit or reworked parts; acquiring representative baseline distributions can be difficult or expensive for COTS devices \cite{trojan_physical_inspection}.

Given the practical realities above, there exists a critical niche for post-silicon, external, and design-agnostic detection methods. These approaches will operate without designer cooperation, without modifying silicon, and with only black-box access to device pins or interfaces. This niche is especially relevant for legacy microcontrollers and COTS devices that dominate many supply chains. External methods can combine non-invasive electrical fingerprinting, side-channel measurements (power, frequency, EM), and functional stress tests designed to exercise rare trigger conditions. By focusing on statistical and side-channel signatures, such methods avoid the cost of deep imaging while retaining applicability across vendor boundaries. Nonetheless, they trade off some detection granularity as they are typically less able to localize the exact inserted circuitry than imaging or design-aware techniques and may require larger sample sizes or more sophisticated signal processing to control false positives \cite{trojan_physical_inspection, trojan_self_testing}.

\subsection{Defensive Side-Channel Anomaly Detection}
Side-channel analysis has historically been viewed as an offensive technique, most notably for extracting cryptographic secrets through differential power analysis and electromagnetic (EM) leakage attacks. A substantial body of early work demonstrated that secret keys and other sensitive parameters could be recovered from unintended physical emissions during computation. More recent research, however, has shown that side-channels can also be leveraged defensively, providing non-intrusive visibility into the internal behavior of embedded systems. In this defensive framing, power and EM emissions are treated as behavioral fingerprints rather than leakage sources, enabling detection of unauthorized code execution, firmware modification, and hardware Trojans \cite{sca3,sca4}. This perspective is particularly relevant for semiconductor supply-chain security, where scalable, non-destructive screening techniques are needed for commercial off-the-shelf components.

Power and EM side-channel signals encode detailed information about microarchitectural activity, instruction sequences, and control-flow transitions. Prior work has demonstrated that distinct firmware routines produce repeatable side-channel signatures, allowing legitimate execution to be fingerprinted and verified. EM-based fingerprinting approaches have shown that deviations in control flow or firmware state can be detected without modifying the target device or embedding additional sensors \cite{sca4,sca5}. Similar concepts have been applied to power measurements, where the execution of specific routines yields characteristic power traces that can serve as baselines for integrity checking \cite{sca3}.

Unlike offensive side-channel attacks, which aim to infer secrets, defensive approaches treat side-channels as out-of-band sensors. By passively monitoring emissions, an external observer can determine whether a device is executing authorized firmware, even when debugging interfaces are disabled or unavailable. This property makes side-channel-based monitoring attractive for high-assurance systems and post-manufacturing inspection in complex global supply chains. A significant body of research has explored EM side-channels for detecting hardware Trojans and malicious modifications. Chip-free EM statistical analysis has been shown to identify anomalous behavior introduced by Trojan circuitry without requiring invasive probing or physical modification of the device \cite{sca4}. Subsequent work demonstrated that EM side-channels can also reveal malicious firmware or runtime malware, particularly when combined with machine-learning classifiers trained on benign execution traces \cite{sca5,sca6}.

Power-based techniques have similarly been used to detect unauthorized code execution. Control-flow tracking via power side-channels demonstrated that power traces could be correlated with instruction-level execution, enabling identification of unexpected code paths \cite{sca3}. More recent efforts have focused on anomaly detection across the device lifecycle, modeling how power signatures evolve over time and across operational modes \cite{sca2}. Notably, several of these studies rely on ChipWhisperer-based measurement setups to collect high-resolution power traces from microcontrollers, illustrating that low-cost, widely available instrumentation can support practical defensive side-channel analysis \cite{sca2,sca3}. Neural-network-based methods have further extended these ideas, particularly for EM side-channels, where learned models are used to classify traces as benign or malicious \cite{sca6,sca7}. While effective in controlled settings, these approaches typically rely on supervised learning and require labeled examples of malicious behavior.

Despite their promise, existing approaches exhibit limitations when viewed from the perspective of scalable supply-chain screening. Many studies emphasize EM measurements, which can be sensitive to probe placement, environmental noise, and laboratory conditions, reducing repeatability outside controlled environments \cite{sca4,sca5}. Power measurements, especially those compatible with ChipWhisperer-style platforms, are often more reproducible and easier to integrate into automated test workflows \cite{sca2}. Additionally, much of the literature focuses on laboratory demonstrations or runtime malware detection rather than explicit supply-chain screening use cases. Screening applications demand high throughput, minimal device modification, and robustness to previously unseen threats. While machine learning is widely used, most prior work relies on supervised classifiers that assume access to representative malicious training data \cite{sca6,sca7}. In realistic screening scenarios, such data may be unavailable or incomplete, limiting generalization. Finally, there has been limited exploration of generative or one-class models for side-channel-based integrity verification. As a result, many existing approaches struggle to detect subtle or novel modifications that fall outside their training distributions.

In contrast to prior work, the present approach treats power side-channels as a foundation for scalable, one-class anomaly detection tailored to semiconductor supply-chain screening. High-fidelity power traces collected using ChipWhisperer-class tools are used to model benign device behavior without requiring examples of specific attacks \cite{sca2,sca3}. Generative models enable learning the distribution of legitimate power signatures and flag deviations indicative of injected code, firmware tampering, or hardware Trojans \cite{sca8,sca10}. This framing explicitly shifts side-channel analysis from an offensive or runtime-monitoring paradigm toward a screening and assurance mechanism. By emphasizing power rather than EM measurements and generative modeling rather than supervised classification, this approach addresses key limitations in prior work and aligns side-channel analysis with the operational constraints of real-world supply-chain security.

Side-channel analysis is no longer limited to secret extraction. Prior work has demonstrated that EM and power side-channels can be used to fingerprint control flow, detect malware, and identify hardware Trojans \cite{sca2,sca3,sca4,sca5,sca6}. However, existing approaches often rely on EM measurements, laboratory-style setups, or supervised learning assumptions that limit scalability. By leveraging power side-channels and generative one-class models, defensive side-channel analysis can evolve into a practical, non-destructive tool for semiconductor supply-chain integrity assurance.

\subsection{GAN-Based Anomaly Detection}
Side-channel measurements such as power consumption and electromagnetic emissions provide a direct view into the runtime behavior of embedded systems. Prior work has demonstrated that power traces encode instruction-level activity, control flow, and firmware-dependent behavior, enabling applications such as code execution tracking and runtime integrity monitoring \cite{sca3, sca2}. Because these signals arise from aggregate switching activity across digital logic, even small changes to firmware or hardware structure can induce measurable deviations in side-channel behavior. From a supply-chain security perspective, this property is particularly valuable. Malicious modifications introduced during fabrication, integration, or firmware provisioning are often designed to evade conventional functional testing by preserving nominal input-output behavior. In realistic screening and validation scenarios, only trusted reference devices or known-good firmware images are available, while the space of possible tampering strategies is unknown and potentially unbounded. This naturally motivates one-class or unsupervised anomaly detection approaches that model normal behavior and flag deviations without requiring labeled examples of malicious activity \cite{sca2}.

Generative Adversarial Networks (GANs) provide a flexible framework for learning complex, high-dimensional data distributions under these constraints. A GAN consists of a generator $G$, which maps samples from a latent distribution to the data space, and a discriminator $D$, which attempts to distinguish real samples from generated ones. Through adversarial training, the generator learns to produce increasingly realistic samples while the discriminator learns a decision boundary separating real data from synthetic approximations \cite{goodfellow2014gan}. Unlike explicit density estimation methods, GANs can capture rich structure in signals such as images and time-series without requiring handcrafted features.

In a one-class anomaly detection setting, GANs are trained exclusively on data representing normal operation. Under these conditions, the discriminator does not learn semantic class boundaries, but instead approximates the support of the normal data distribution. Samples outside this learned support receive lower discriminator confidence and can be flagged as anomalous \cite{zenati2018efficient}. This interpretation has been formalized and validated across multiple domains, including vision, medical imaging, and time-series analysis, where anomalous examples are rare or unavailable during training \cite{schlegl2017unsupervised, akcay2019ganomaly}.

Several strategies have been proposed for performing anomaly detection with GANs. One common approach uses the discriminator output directly as a normality score. Because the discriminator is optimized to separate real training samples from generated ones, it implicitly learns the manifold of normal behavior, and deviations from this manifold result in reduced discriminator confidence \cite{zenati2018efficient}. This approach is computationally efficient and well suited to high-throughput or real-time applications. An alternative strategy relies on reconstruction or projection error. In these methods, an input sample is mapped into the generator’s latent space, and the discrepancy between the original sample and its reconstruction is measured. Samples that cannot be accurately reconstructed are assumed to lie outside the normal data manifold \cite{schlegl2017unsupervised}. Variants such as GANomaly and OCGAN combine reconstruction error with discriminator-based features or impose constraints on the latent space to improve separation between normal and abnormal samples \cite{akcay2019ganomaly, perera2019ocgan}. While effective, these methods typically incur higher inference cost due to iterative optimization or additional network components.

Power side-channel traces are a natural fit for one-class GAN-based modeling. Past work has shown that power measurements can be used to fingerprint firmware versions, detect control-flow deviations, and identify anomalous execution behavior in embedded systems \cite{sca3, sca2}. Unlike software instrumentation or on-chip sensors, external power measurement is non-invasive and does not require design-time cooperation, making it applicable to commercial off-the-shelf components and legacy devices.


\section{SYSTEM AND THREAT MODEL}

\label{sec:model}

\subsection{System Model}
The proposed screening system consists of three components: a device under test (DUT) executing a standardized workload, a power measurement apparatus that captures side-channel traces, and a one-class anomaly detection model based on generative adversarial networks. The system operates in two phases. During the enrollment phase, power traces are collected from a population of authenticated benign devices and used to train a GAN. During the screening phase, power traces from incoming devices are scored by the trained discriminator, and devices exhibiting anomalous scores are flagged for further investigation.

The DUT is a commercial off-the-shelf microcontroller or embedded device that executes a fixed cryptographic or computational workload. The workload must be repeatable and produce consistent control flow across devices to enable meaningful comparison of power signatures. No modifications to the device silicon, firmware, or package are required, making the approach applicable to COTS components where design access is unavailable. Power measurements are collected via external current sensing, preserving device integrity and enabling screened units to be deployed without concerns about warranty voiding or induced defects.

The defender controls the test environment, including workload selection, measurement parameters (sampling rate, window duration, trigger configuration), preprocessing steps (cropping, normalization), and model training hyperparameters. The defender also calibrates detection thresholds based on the desired false positive rate, enabling tuning of the sensitivity-specificity tradeoff according to available forensic analysis capacity. Screening does not require access to trusted manufacturer cooperation, reflecting realistic constraints in downstream supply chain scenarios \cite{s20185165}.

\begin{figure}
    \centering
    \includegraphics[width=1\linewidth]{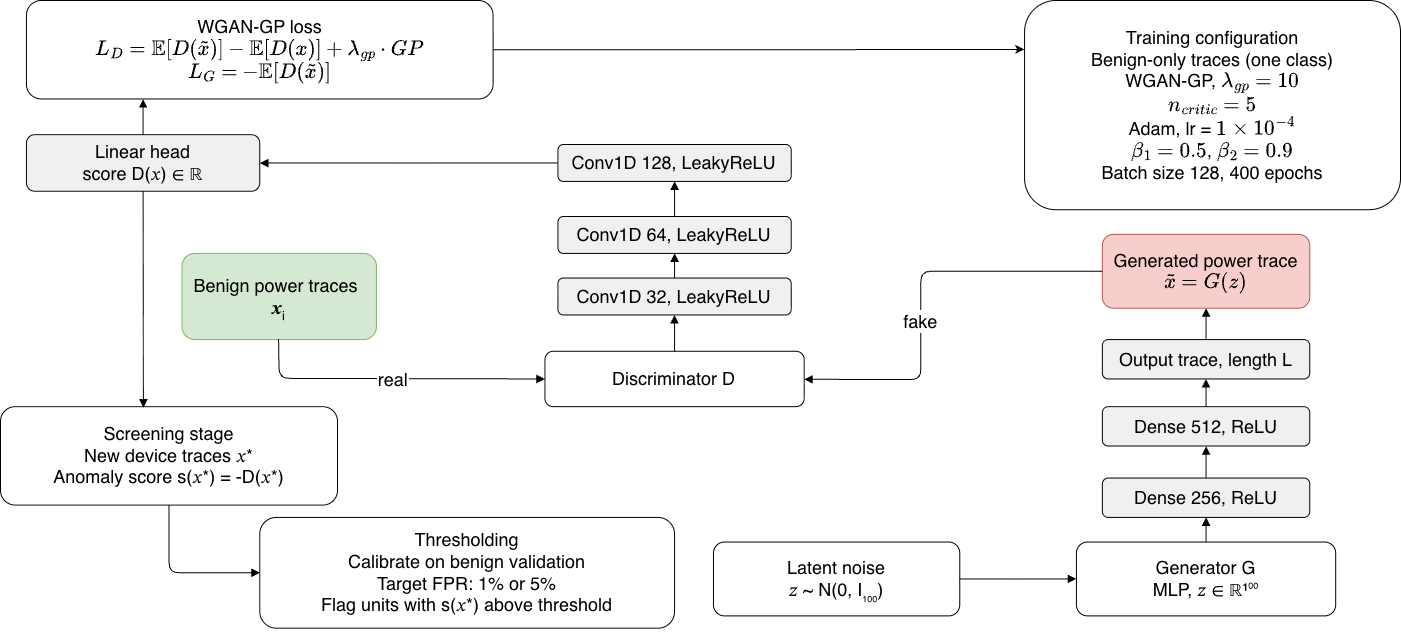}
    \caption{System Model Overview}
    \label{fig:sysDiag}
\end{figure}

An overview of this workflow is featured in algorithm \ref{alg:screening} and figure \ref{fig:sysDiag}.

\begin{algorithm}[h]
\caption{Power Side-Channel Screening Workflow}
\label{alg:screening}
\begin{algorithmic}[1]
\Require Benign trace set $\mathcal{B}$, incoming device $d$, discriminator $D$, thresholds $\tau_{1\%}, \tau_{5\%}$
\Function{Enroll}{}
    \State Collect traces $\mathcal{B}$ from authenticated devices under workload $W$
    \State Compute global mean $\mu$ and std.\ deviation $\sigma$ of $\mathcal{B}$
    \State Train WGAN-GP on normalized $\mathcal{B}$, store discriminator $D$
    \State Calibrate thresholds $\tau_{1\%}, \tau_{5\%}$ using validation split of $\mathcal{B}$
\EndFunction
\Function{Screen}{device $d$, operating point $\tau$}
    \State Acquire normalized trace $x$ from $d$ under workload $W$
    \State Score $s = -D(x)$
    \If{$s \ge \tau$}
        \State Flag $d$ for quarantine and downstream forensic inspection
    \Else
        \State Approve $d$ for integration
    \EndIf
\EndFunction
\end{algorithmic}
\end{algorithm}

\subsection{Threat Model}
The adversary is assumed to possess the capability to modify device firmware or insert hardware Trojans at any point upstream of the screening facility. The objective is to introduce malicious functionality that remains dormant during standard functional testing and activates only under specific post-deployment trigger conditions. Such triggers may include rare input patterns, cryptographic subversion mechanisms, or deliberate manipulation of shared resources.

The threat model further assumes familiarity with the target microcontroller architecture and a high-level understanding of the screening workload, for example, awareness that AES encryption is executed. Access to the specific plaintext inputs used during screening, the exact timing of the measurement window, and the trained model parameters and decision thresholds is not available to the adversary. The measurement infrastructure, including signal acquisition, preprocessing, and threshold calibration, is assumed to be fully controlled by the defender and cannot be manipulated by the attacker.

Detection is explicitly limited to adversarial modifications that induce observable deviations in power consumption under the standardized workload. Attacks that operate below the measurement noise floor, activate only under environmental conditions absent from the test fixture, or alter functionality without measurably affecting power signatures may evade detection. These constraints position the proposed approach as a complementary assurance mechanism, intended to augment rather than replace broader multi-modal hardware and firmware validation techniques.

\subsection{Security Goals}
The primary objective is to identify devices whose power consumption profiles deviate statistically from the learned benign baseline while maintaining low false positive rates. Anomalous devices are quarantined and escalated to high-assurance forensic methods such as X-ray imaging or destructive analysis. The screening must be non-destructive, operate without modifying devices, and achieve throughput compatible with incoming inspection workflows.

The one-class training approach eliminates the need to enumerate all possible tampering variants. The model learns the manifold of expected benign behavior, and any deviation, whether from firmware tampering, hardware modification, or counterfeit substitution, is treated as evidence of potential compromise. Detection thresholds are set to achieve target false positive rates of 1 to 5\%, ensuring most legitimate devices pass screening while a small fraction is selected for deeper inspection. This enables efficient allocation of expensive forensic resources across a large number of microcontrollers.

\subsection{Deployment Points In The Supply Chain}
\subsubsection{End-of-Line Manufacturing and Final Test}
At the manufacturing stage, screening can be integrated immediately following packaging and final electrical test, when devices have stable interfaces and are accessible for measurement \cite{john2015amkor_testflow}. This upstream deployment point offers the earliest opportunity to establish component integrity before distribution. Screening can be performed on a per-lot sampling basis or, where throughput permits, on every unit prior to shipment \cite{dfars2522467007}. While this placement provides the broadest coverage, it typically requires coordination with manufacturers or authorized distributors and may not be accessible to downstream integrators acquiring components through standard commercial channels \cite{dimase2016traceability_risk}.

\subsubsection{Receiving Inspection and Board Assembly}
The most operationally viable deployment point for many organizations is at receiving inspection, where integrators first take custody of purchased components \cite{dfars2522467007, dimase2016traceability_risk}. At this stage, the screening fixture integrates naturally with existing acceptance testing infrastructure. Standard functional verification and boundary scan can serve as a first-pass filter to reject grossly defective parts, after which power-based screening provides a second-tier integrity check for devices that pass basic I/O validation \cite{maunder1991_ieee1149_1, cathis2024sok_power_malware}.

This two-stage approach optimizes resource allocation as the majority of devices that exhibit both functional correctness and normal power signatures can be cleared for assembly with high confidence, while the small fraction flagged as anomalous by the discriminator can be quarantined for deeper investigation \cite{cathis2024sok_power_malware}. Depending on risk tolerance and available forensic capacity, these flagged devices may undergo X-ray imaging, decapsulation, or destructive reverse engineering \cite{doe2024_counterfeit_handbook}. This tiered strategy enables integrators to achieve substantially higher assurance coverage than would be possible if every device required expensive forensic analysis, while still directing those resources toward the units most likely to warrant scrutiny \cite{doe2024_counterfeit_handbook}.

\subsubsection{Depot, Refurbishment, and Sustainment}
Screening at depot enables continuous assurance rather than a one-time receiving gate \cite{sca2}. Components can be re-screened during firmware updates, board-level repairs, or component replacement cycles \cite{sca2}. For systems with long lifespans, this repeated verification capability provides ongoing visibility into component integrity throughout the maintenance lifecycle \cite{cathis2024sok_power_malware}. Screening can be performed either at the board level when sufficient signal isolation is achievable, or at the component level during reprogramming or replacement procedures \cite{cathis2024sok_power_malware}.

\subsubsection{Operational Integration and Model Management}

Across all deployment points, the screening architecture supports a consistent operational model. Detection thresholds are calibrated to achieve target false positive rates, typically 1\% to 5\%, based on available downstream forensic capacity and acceptable inspection overhead \cite{sca2, akcay2019ganomaly}. Organizations with greater forensic resources can afford higher false positive rates to maximize detection sensitivity, while resource-constrained facilities can set more conservative thresholds to reduce the volume of escalated cases \cite{dfars2522467007}.

The approach does require disciplined workload standardization and model lifecycle management. Screening models are device-specific and workload-dependent, as a model trained on one microcontroller executing AES encryption will not generalize to a different chip architecture or a different computational workload \cite{cathis2024sok_power_malware, sca2}. Furthermore, changes to firmware versions, compiler toolchains, clock configurations, or voltage domains may alter power signatures sufficiently to require model retraining \cite{cathis2024sok_power_malware, sca2}.

In practice, this implies maintaining an enrollment pipeline as part of configuration management. When a new device variant is introduced, when firmware baselines are updated, or when measurement infrastructure changes, a new enrollment phase must be executed to collect authenticated benign traces and retrain the discriminator \cite{cathis2024sok_power_malware}. Periodic recalibration may also be necessary to account for environmental drift, fixture aging, or shifts in manufacturing process that affect nominal power characteristics \cite{cathis2024sok_power_malware, sca2}.

When managed systematically, however, this model lifecycle overhead is no greater than the version control and calibration requirements already present in standard semiconductor test and inspection workflows \cite{john2015amkor_testflow}. The screening station can be integrated into existing test sequences as an automated decision aid, producing a quantitative anomaly score that informs risk-based escalation decisions without requiring manual interpretation of raw power waveforms \cite{cathis2024sok_power_malware}.

\subsubsection{Positioning Within Tiered Assurance Strategy}
The power-based screening tier bridges a critical gap between two extremes in current practice. Functional verification and basic electrical testing are fast and inexpensive, but provide negligible assurance against sophisticated modificat that preserves nominal input-output behavior \cite{cathis2024sok_power_malware}. Conversely, high-assurance forensic techniques such as X-ray tomography and destructive delayering provide strong evidence of hardware modifications, but are too slow and costly to apply comprehensively \cite{doe2024_counterfeit_handbook}.

By producing a rapid, quantitative integrity signal for every screened device, the proposed approach enables efficient allocation of expensive forensic resources \cite{dfars2522467007, doe2024_counterfeit_handbook}. Instead of blind sampling or subjective judgment, forensic analysis can be directed toward devices exhibiting anomalous power behavior \cite{cathis2024sok_power_malware}. This data-driven prioritization increases the likelihood that forensic investment is applied to genuinely compromised components while avoiding unnecessary inspection of legitimate devices \cite{dfars2522467007}.

The result is a scalable assurance architecture in which most devices pass quickly through automated screening, a small fraction are flagged for human review or deeper testing, and an even smaller subset undergo full forensic workup \cite{doe2024_counterfeit_handbook}. This tiered strategy achieves substantially higher effective coverage than would be possible with forensic methods alone, while maintaining throughput compatible with operational inspection and sustainment timelines \cite{dfars2522467007, doe2024_counterfeit_handbook}.

\section{Experimental Methodology}\label{sec:experimental}

\subsection{Hardware and Test Setup}
The Atmel XMEGA128D4 (XMEGA) microcontroller is selected as the target for this experiment as a representative example of auxiliary microcontrollers that are routinely integrated into embedded and cyber-physical systems. These devices often retain direct control over safety-relevant or system-critical functions despite not serving as the primary processing element of the system. Their supporting role frequently places them outside the scope of detailed forensic assessment and post-incident analysis, which typically focuses on higher-level processors and application software \cite{Kovar2018} This dynamic creates a condition in which compromise of a seemingly peripheral microcontroller can induce system-level failures, making such devices attractive targets for cyber-physical attacks.

\begin{figure}
    \centering
    \includegraphics[width=1\linewidth]{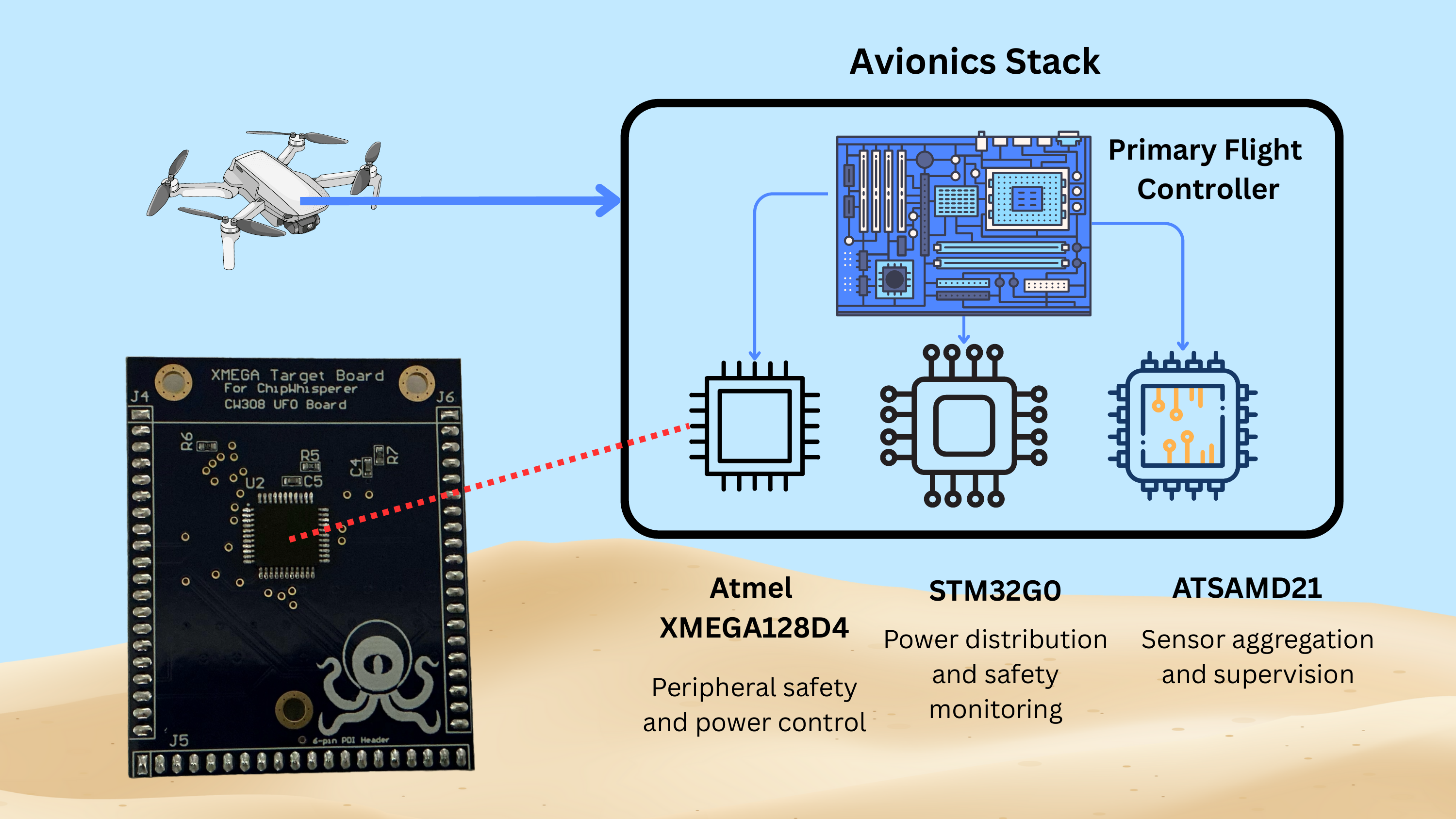}
    \caption{Representative avionics stack illustrating the role of tertiary microcontrollers in a COTS unmanned aerial system. The Atmel XMEGA128D4, STM32G0, and ATSAMD21 are examples of peripheral controllers that typically receive limited forensic scrutiny \cite{Kovar2018}, despite exercising authority over safety-relevant functions and constituting potential single points of failure.}
    \label{fig:drones}
\end{figure}

An example of this architectural pattern can be found in a COTS drone or UAS platform. The XMEGA may be employed as a dedicated peripheral management or safety controller, while a separate, higher-performance processor executes navigation and control algorithms. The XMEGA would be tasked with functions such as sensor polling, motor enable and arming logic, battery voltage and current monitoring, and enforcement of hardware-level safety thresholds. Manipulation of this microcontroller could result in falsified sensor readings, bypassed safety interlocks, or deliberate disruption of actuator control, even though the primary control processor continues to operate nominally. This data flow is depicted in figure \ref{fig:drones}.

\begin{figure}
    \centering
    \includegraphics[width=1\linewidth]{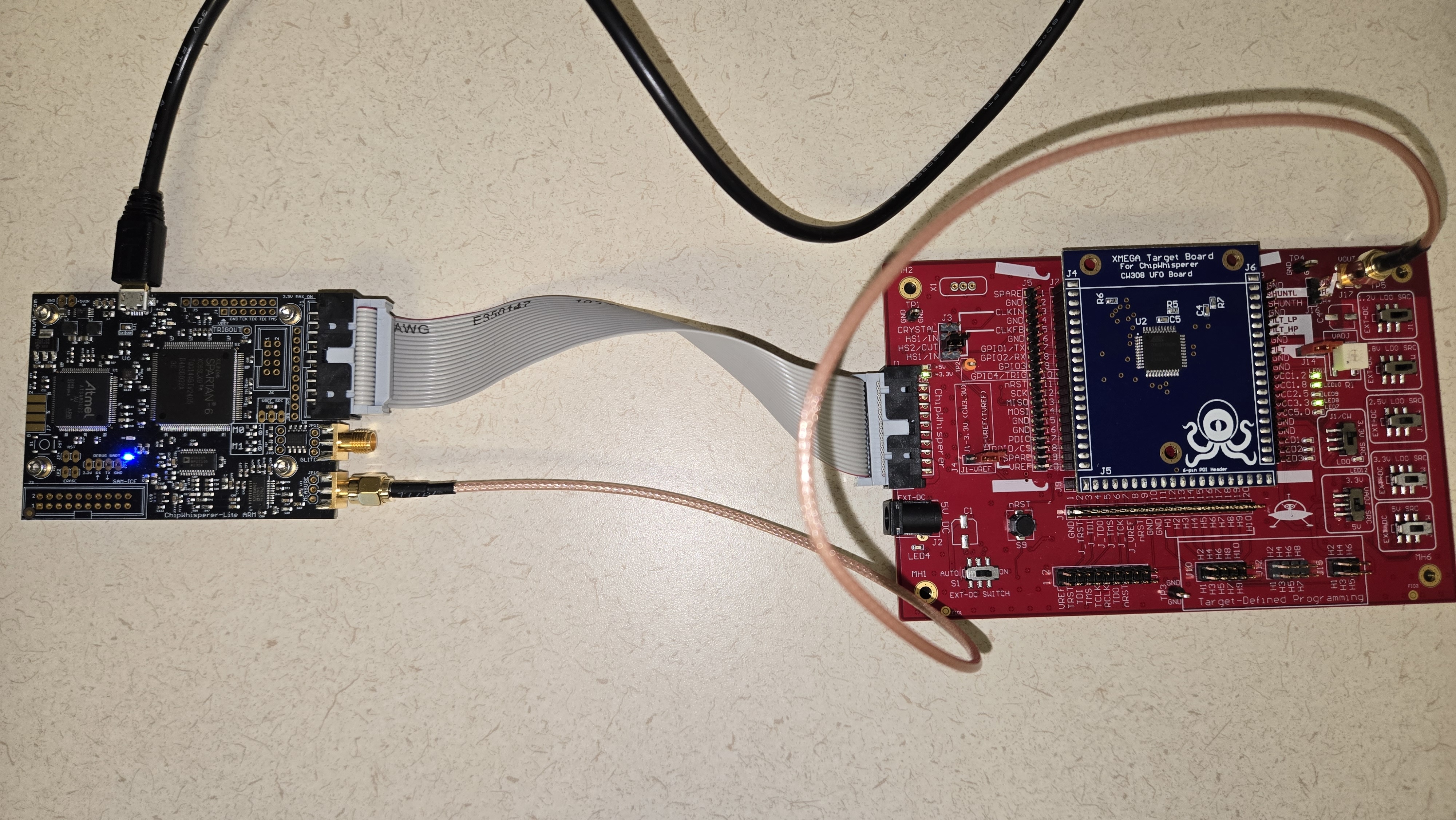}
    \caption{Experimental power side-channel measurement setup. A ChipWhisperer-Lite (CW1173) capture board is interfaced with a CW308 UFO baseboard hosting an Atmel XMEGA128D4 target microcontroller. The CW308 platform exposes the shunt-based power measurement point and SimpleSerial interface, while the ChipWhisperer provides synchronized clock injection, triggering, and trace acquisition for repeatable power side-channel analysis.}
    \label{fig:chipwhisperersetup}
\end{figure}

Power traces are collected using a ChipWhisperer-Lite (CW1173) capture board interfaced with a CW308 UFO baseboard hosting the Atmel XMEGA128D4 target microcontroller. This setup is depicted in figure \ref{fig:chipwhisperersetup}.

The ChipWhisperer-Lite provides synchronized trace acquisition and clock injection, while the CW308 platform exposes the target’s shunt-based power measurement point and SimpleSerial interface for repeatable workload execution and trigger alignment.

Power side-channel measurements are acquired across the on-board shunt resistor and digitized at 10-bit resolution. Each acquisition records 5,000 samples, which are cropped to a 3,000-sample analysis window centered on the AES computation. Trace acquisition is synchronized using the default SimpleSerial GPIO1 trigger, asserted by the target firmware at the start of the AES routine to ensure temporal alignment across captures.

The analog front-end gain is configured for the 0 to 3.3 V supply domain. The target is clocked using the ChipWhisperer clock generator routed through the HS2 output, and power is supplied directly by the CW-Lite platform. All scope parameters follow the ChipWhisperer-Lite default configuration to support reproducibility and compatibility with standard capture workflows.

\subsection{Workloads and Tampering Scenarios}
The baseline workload executes the standard ChipWhisperer SimpleSerial AES firmware compiled for the CWLITEXMEGA target platform. Each measurement sequence follows the SimpleSerial v1.0 protocol: a 16-byte plaintext is transmitted to the target via the \texttt{p} command, the microcontroller performs AES-128 encryption using a pre-loaded cryptographic key, and the resulting ciphertext is returned through the \texttt{r} command. This configuration establishes the nominal operational baseline against which anomalous behavior is evaluated.

To assess detection performance across multiple threat modalities, we constructed a composite tampered firmware image incorporating four distinct classes of malicious modification. First, a conditional hardware Trojan is implemented that activates only when the first plaintext byte equals \texttt{0x42}, executing an additional key-dependent computational loop. This represents trigger-based payloads designed to remain dormant during standard functional verification. Second, a hidden backdoor command is introduced through an auxiliary SimpleSerial command (\texttt{b}) that returns a fixed 16-byte secret value. This backdoor path is not exercised during the profiled AES workload and therefore serves as a negative control for workload-bound power side-channel screening. Third, a cryptographic integrity violation is implemented by applying a single-bit flip to each output ciphertext, simulating subtle corruptions such as malformed S-box lookups or incomplete cipher rounds that degrade cryptographic security while maintaining structural output validity. Fourth, a fixed-iteration delay loop is appended following the AES computation to extend execution latency and elevate instantaneous power consumption, consistent with unauthorized data exfiltration routines or resource utilization.

Both the benign and tampered firmware variants were compiled from identical source repositories using the same toolchain configuration and programmed to the target microcontroller via the standard XMEGA flashing procedure. This experimental design ensures that observed divergences in power consumption profiles are attributable to the injected malicious behaviors rather than incidental variation introduced by compilation parameters, programming artifacts, or environmental conditions. Critically, the tampered firmware maintains functional equivalence with respect to standard test vectors, producing cryptographically correct outputs for nominal inputs and thereby passing conventional acceptance testing protocols while exhibiting measurable anomalies in the power side-channel domain.

\subsection{Data Collection and Preprocessing}
We collected 2,000 power traces from the benign firmware and 2,000 traces from the composite tampered firmware, with each trace comprising 3,000 temporal samples stored as 32-bit floating-point values. To isolate the contribution of individual payloads, we additionally captured 1,000 traces for each selective tampered build (conditional Trojan, ciphertext bit flip, delay loop, and hidden backdoor command) by recompiling the firmware so that only one malicious mechanism remained active. The preprocessing pipeline applies temporal cropping to extract the interval from sample 500 to sample 2000, isolating the core AES computation window while excluding initialization overhead and post-processing operations.

Normalization is performed by computing the global mean and standard deviation across all benign training traces, with these statistics subsequently applied to normalize both the benign validation set and the tampered test sets. This approach ensures that the model learns intrinsic distributional properties of benign execution rather than artifacts of scale or offset. The benign dataset is partitioned into an 80/20 training-validation split, with 1,600 traces allocated for model training and 400 traces reserved for threshold calibration and hyperparameter validation. All tampered traces, both composite and per-attack, are withheld entirely from the training process and used exclusively for inference-time evaluation of discriminator detection performance.

\subsection{GAN Model and Training Configuration}
We employ a one-class Wasserstein GAN with gradient penalty (WGAN-GP) trained exclusively on benign power traces. The generator network $G$ maps a 100-dimensional latent vector $z$ sampled from a standard normal distribution to a synthetic power trace of length $L$. The generator architecture consists of two fully connected hidden layers with 256 and 512 units respectively, each followed by ReLU activation, culminating in a linear output layer that is reshaped to produce a one-dimensional trace of shape $(1, L)$.

The discriminator network $D$ implements a one-dimensional convolutional critic architecture with three successive convolutional blocks employing 32, 64, and 128 filters respectively, each followed by LeakyReLU activation with a negative slope of 0.2. The convolutional feature maps are flattened and processed through a final linear layer that outputs a scalar critic score $D(x)$ representing the Wasserstein distance estimate for input trace $x$.

Training follows the WGAN-GP objective with a gradient penalty coefficient $\lambda_{gp} = 10$ and $n_{critic} = 5$, meaning the discriminator is updated five times for each generator update to maintain training stability. Both networks are optimized using the Adam optimizer with a learning rate of $1\times10^{-4}$ and momentum parameters $\beta_1 = 0.5$ and $\beta_2 = 0.9$. Training proceeds for 400 epochs with a batch size of 128 samples. Only benign traces are used during training and validation phases to enforce the one-class learning paradigm.

At inference time, new traces $x^*$ are processed through the trained discriminator $D$, and an anomaly score is computed as $s(x^*) = -D(x^*)$, where higher scores indicate greater deviation from the learned benign distribution. Detection thresholds are calibrated on the benign validation set to achieve target false positive rates, such as 1\% or 5\%, enabling flexible adjustment of the sensitivity-specificity tradeoff based on operational requirements and downstream inspection capacity.

\subsection{Baselines and Anomaly Scoring}
The primary anomaly scoring mechanism utilizes the negated discriminator output, where lower (more negative) critic scores correspond to traces that deviate more substantially from the learned benign manifold. Threshold values are empirically determined by sorting benign validation scores and selecting cutoff points that achieve the desired false positive rates of 1\% and 5\%. This calibration approach ensures that the screening system can be tuned to match available forensic analysis capacity while maintaining consistent error rates across deployment contexts. 

\section{RESULTS}
\label{sec:results}

\subsection{Detection Performance Across Scenarios}
The discriminator was trained on 1,600 benign power traces with 400 additional benign traces reserved for validation and threshold calibration. Detection performance was evaluated on 2,000 tampered traces collected from the composite malicious firmware. Using the negated discriminator score as the anomaly metric, the model achieves a receiver operating characteristic area under curve (ROC AUC) of 0.995, indicating strong separation between benign and tampered distributions. At a false positive rate of 1\%, the true positive rate is 0.932, meaning 93.2\% of tampered devices are correctly flagged while only 1\% of benign devices trigger false alarms. At a 5\% false positive rate, the true positive rate increases to 0.971, capturing 97.1\% of tampered devices.

Table~\ref{tab:data-summary} summarizes the characteristics of the captured datasets. All traces consist of 3,000 temporal samples with no missing or invalid values. 

The composite tampered firmware produced measurable perturbations across multiple dimensions of the power signature. Conditional execution paths resulted in intermittent bursts of elevated power consumption, creating temporal variability not present in the benign baseline. The introduction of auxiliary protocol handling altered timing characteristics and shifted the distribution of instantaneous current draw during communication phases. Modifications to cryptographic output processing introduced subtle changes to downstream computation patterns, affecting the statistical properties of power consumption in later portions of the trace. The addition of post-computation delay loops consistently elevated average power and extended execution time, producing a sustained deviation from nominal energy profiles. Figures~\ref{fig:roc} and \ref{fig:score-hist} plots the ROC curve alongside the benign and tampered score histograms that substantiate these quantitative claims.

\begin{figure}[H]
    \centering
    \includegraphics[width=0.75\linewidth]{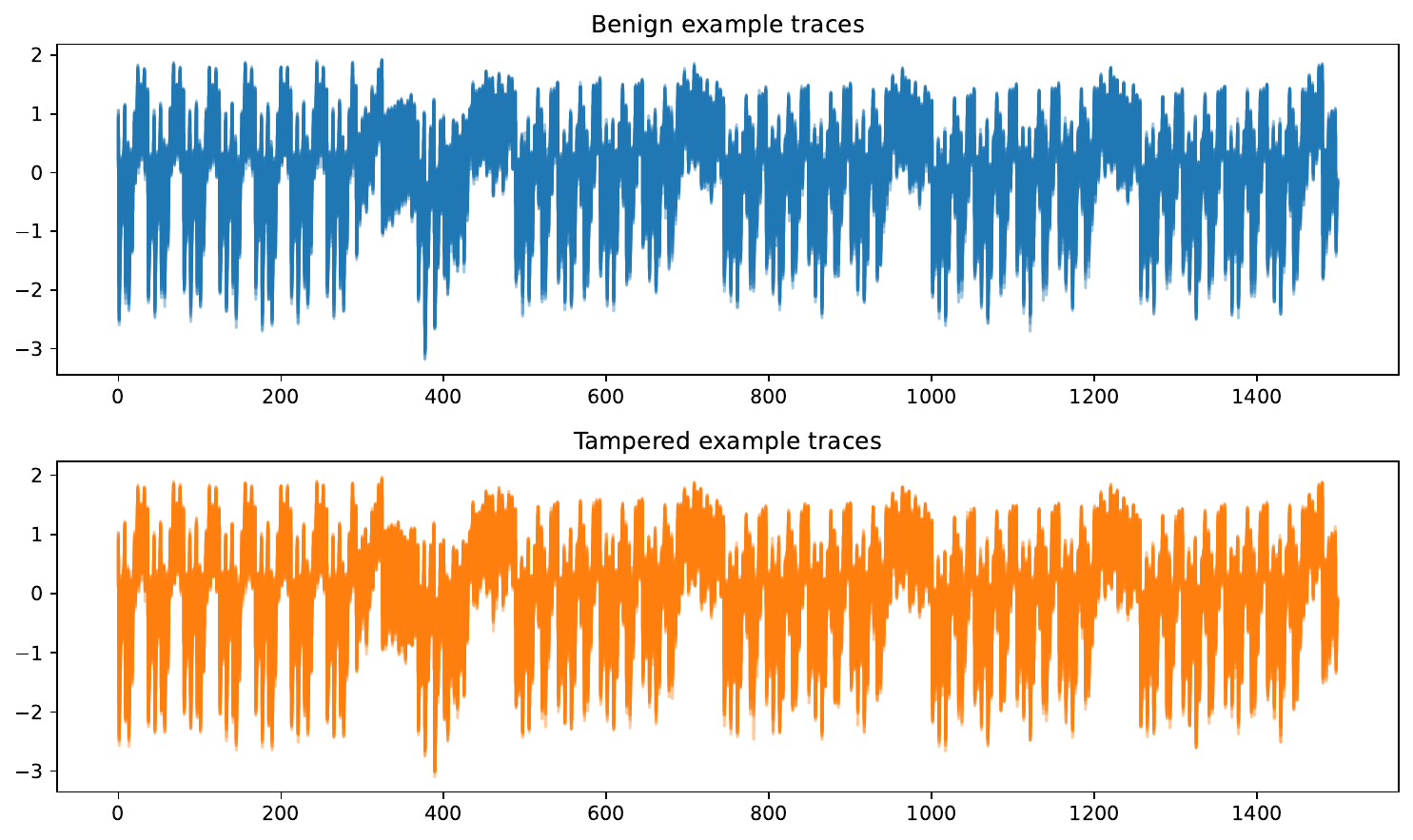}
    \caption{Overlay of representative benign and tampered power traces highlighting temporal deviations induced by malicious modifications.}
    \label{fig:trace-overlays}
\end{figure}

\begin{figure}[H]
    \centering
    \includegraphics[width=0.75\linewidth]{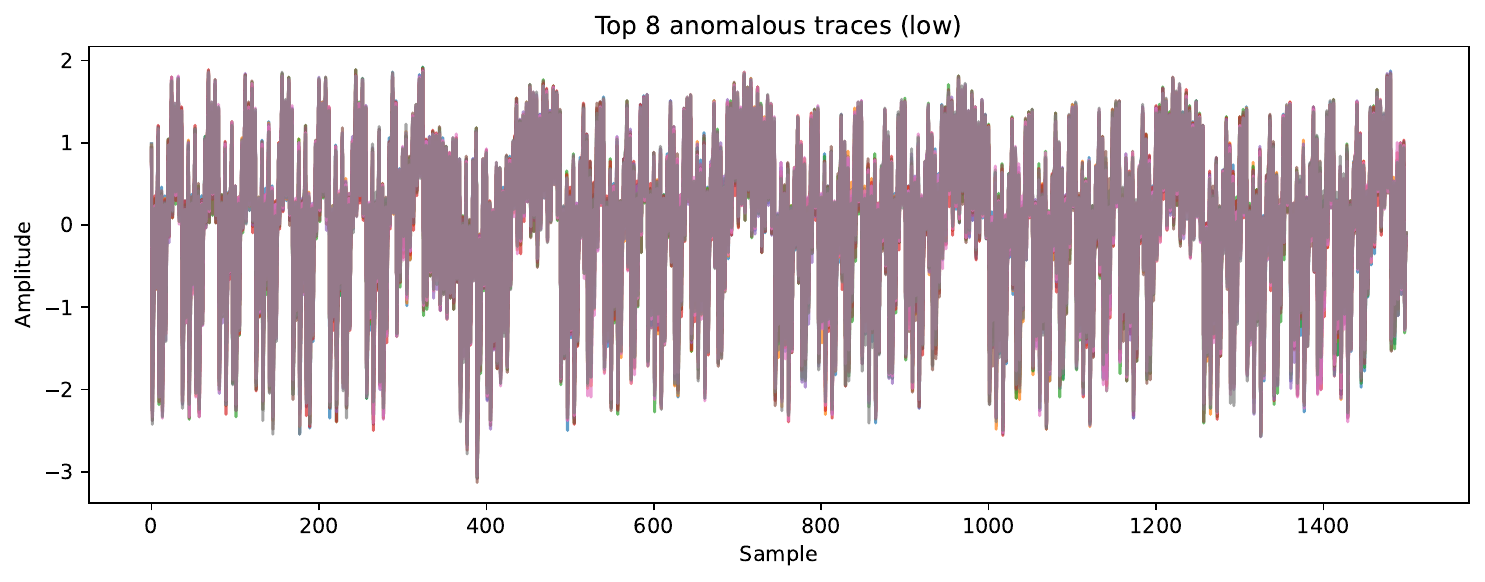}
    \caption{Most anomalous tampered power traces ranked by discriminator score.}
    \label{fig:top-traces}
\end{figure}

These perturbations are reflected in the strong separation of discriminator score distributions. Tampered traces consistently receive higher anomaly scores than benign traces, with median scores of approximately 9.70 for tampered versus 9.41 for benign. The discriminator successfully learned to recognize these deviations despite training exclusively on benign examples, demonstrating that the one-class formulation can generalize to detect diverse tampering modalities without requiring prior knowledge of specific attack implementations. Figures~\ref{fig:trace-overlays} and \ref{fig:top-traces} highlight representative waveform overlays and the most anomalous tampered traces to illustrate the temporal structure of these deviations.
\begin{table}[h]
\centering
\caption{Captured datasets (raw traces, single device/session).}
\label{tab:data-summary}
\begin{tabular}{lcccccc}
\hline
Class & Traces & Samples & Min & Max & Mean & Std \\\hline
Benign & 2{,}000 & 3{,}000 & $-0.253$ & $0.118$ & $-0.013$ & $0.067$ \\
Tampered & 2{,}000 & 3{,}000 & $-0.241$ & $0.118$ & $-0.012$ & $0.067$ \\\hline
\end{tabular}
\end{table}

\begin{table}[h]
\centering
\caption{Detection metrics for the combined tampered firmware (single device/session).}
\label{tab:metrics-summary}
\begin{tabular}{lccccc}
\hline
Scenario & AUC & TPR@1\% FPR & TPR@5\% FPR & Thr@1\% FPR & Thr@5\% FPR \\\hline
Tampered\_aes & 0.995 & 0.932 & 0.971 & $9.566$ & $9.534$ \\\hline
\end{tabular}
\end{table}

To attribute these aggregate results to individual threat mechanisms, we reprogrammed the XMEGA into 4 selective tampered configurations that enable only one payload at a time (rare-trigger Trojan, ciphertext bit flip, delay loop, or hidden backdoor command) and collected 1{,}000 traces for each variant using the same capture setup. Table~\ref{tab:per-attack-metrics} reports discriminator performance per attack. The control-flow heavy Trojan and delay loop workloads remain trivially separable even at a 1\% false positive rate, while the subtle ciphertext perturbation still achieves a defensible 0.94 AUC with 0.53/0.70 TPR at 1\%/5\% FPR. The backdoor command constitutes a negative control for workload-bound screening. In this configuration, malicious functionality is reachable only through an auxiliary SimpleSerial command path that does not execute during the profiled AES workload. Consequently, the power traces captured under the AES-triggered acquisition window remain statistically indistinguishable from benign traces, and the discriminator yields chance-level performance. This result is expected, and it highlights that side-channel screening provides assurance only for code paths exercised by the standardized workload.

\begin{table}[h]
\centering
\caption{Per-attack detection performance (1,000 traces per scenario, single device/session). Backdoor command is a negative control.}
\label{tab:per-attack-metrics}
\begin{tabular}{lccc}
\hline
Scenario & AUC & TPR@1\% FPR & TPR@5\% FPR \\\hline
Conditional Trojan & 1.000 & 1.000 & 1.000 \\
Ciphertext bit flip & 0.944 & 0.526 & 0.701 \\
Delay loop insertion & 0.9998 & 0.998 & 0.998 \\
Backdoor command & 0.000 & 0.000 & 0.000 \\\hline
\end{tabular}
\end{table}

\subsection{Score Distributions And Feature Space Structure}
\begin{figure}[t]
    \centering
    \includegraphics[width=0.5\linewidth]{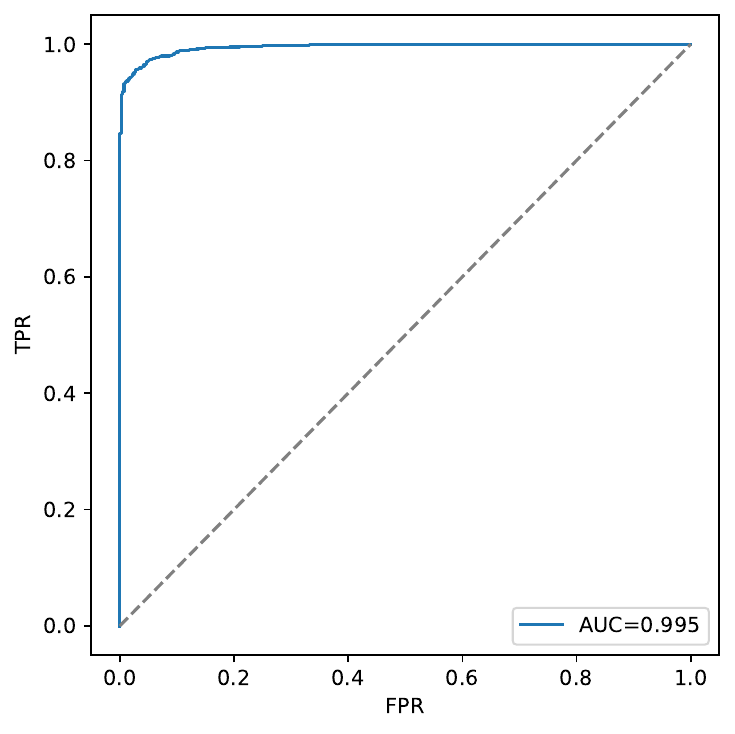}
    \caption{Receiver operating characteristic (ROC) curve for benign versus tampered power traces under the AES workload.}
    \label{fig:roc}
\end{figure}

\begin{figure}[t]
    \centering
    \includegraphics[width=0.5\linewidth]{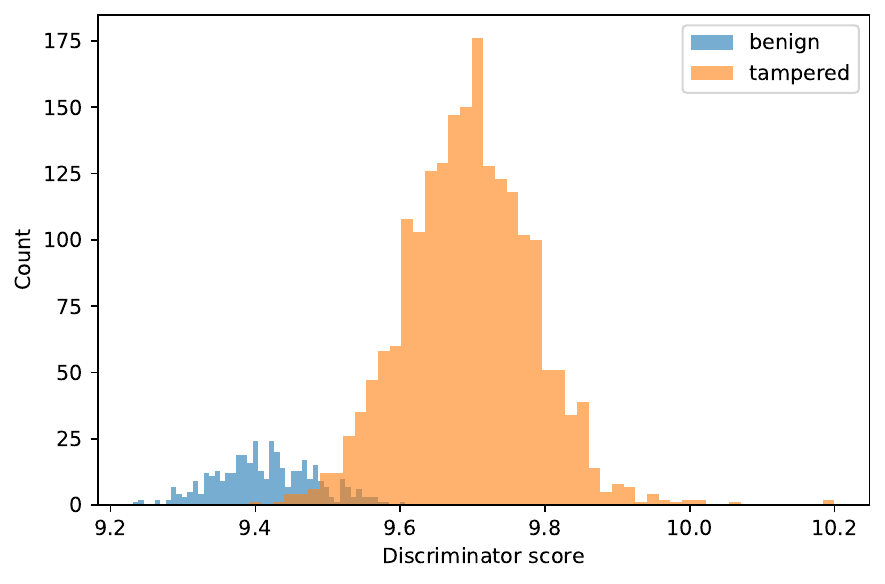}
    \caption{Discriminator score distributions for benign and tampered power traces.}
    \label{fig:score-hist}
\end{figure}

\begin{figure}[t]
    \centering
    \includegraphics[width=0.75\linewidth]{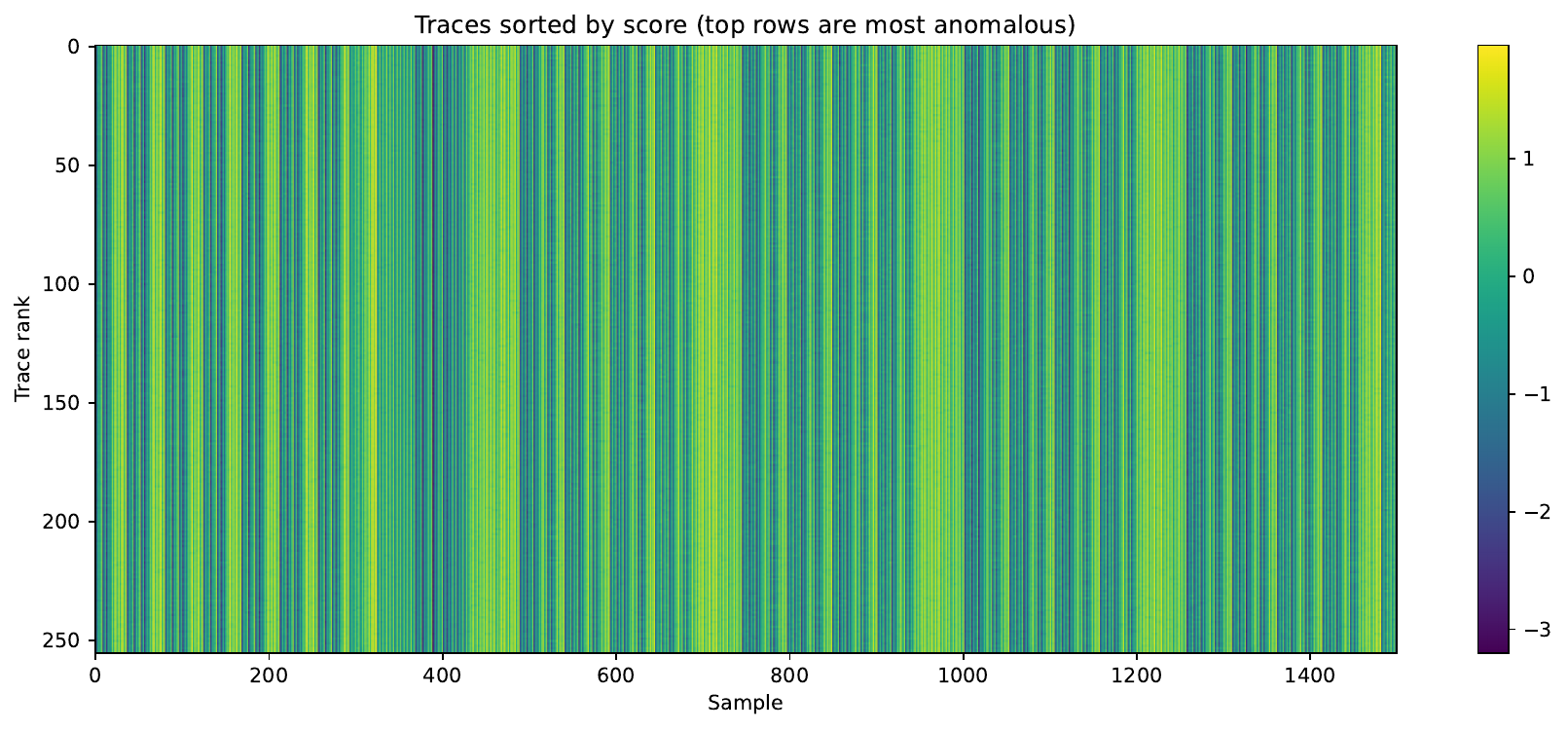}
    \caption{Trace-score ordering heatmap illustrating separation between benign and tampered executions.}
    \label{fig:trace-score-heatmap}
\end{figure}

\begin{figure}[t]
    \centering
    \includegraphics[width=0.75\linewidth]{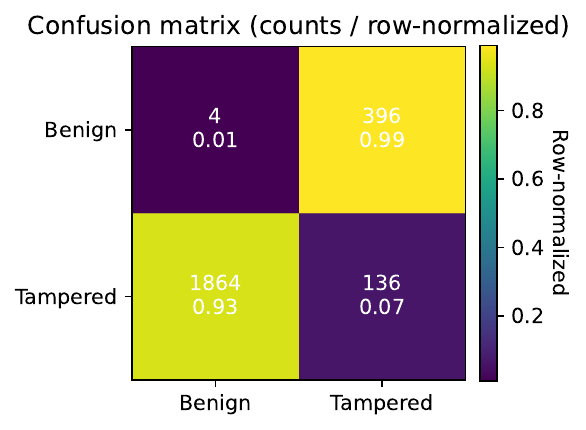}
    \caption{Confusion matrix at the selected operating threshold corresponding to a 1\% false positive rate.}
    \label{fig:confusion}
\end{figure}

\begin{figure}[t]
    \centering
    \includegraphics[width=0.5\linewidth]{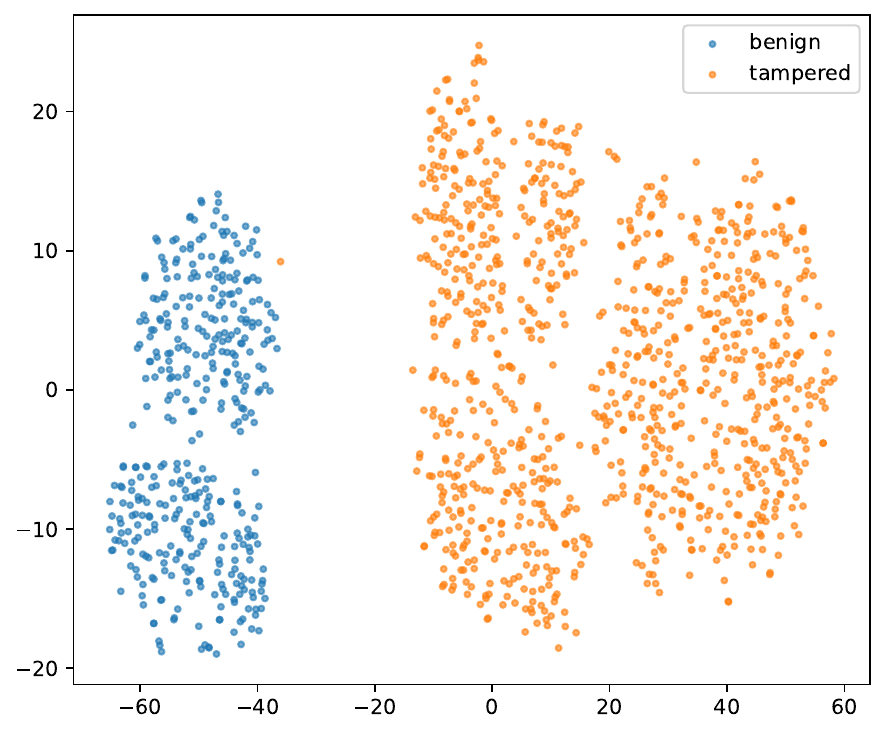}
    \caption{Low-dimensional embedding of discriminator activations showing separation between benign and tampered traces.}
    \label{fig:embedding}
\end{figure}

\begin{figure}[t]
    \centering
    \includegraphics[width=0.5\linewidth]{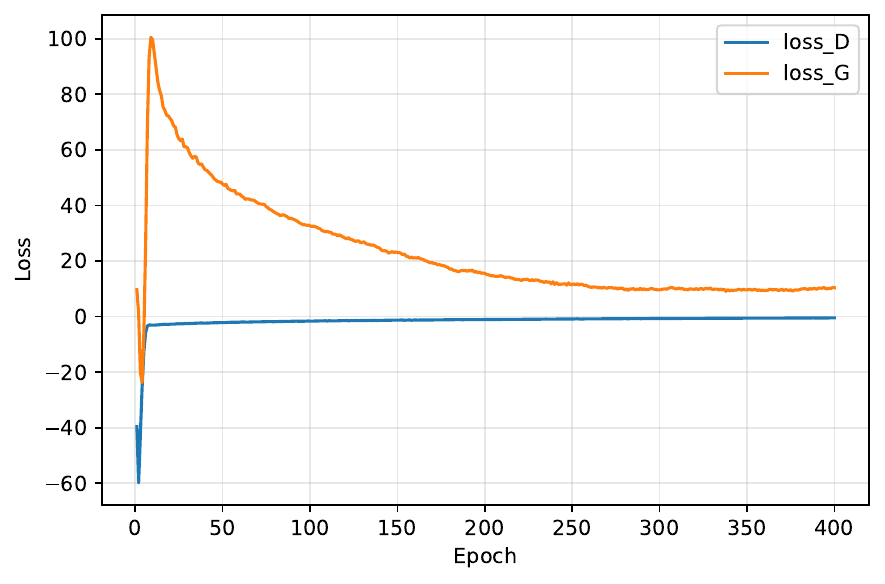}
    \caption{Generator and discriminator training curves demonstrating stable WGAN-GP convergence.}
    \label{fig:training-curves}
\end{figure}

Discriminator scores separate well after sign correction with median benign score $\approx 9.41$ and tampered $\approx 9.70$ with little overlap, as visualized in Figure~\ref{fig:roc} and Figure~\ref{fig:score-hist}. The same operating point yields the confusion matrix shown in Figure~\ref{fig:trace-score-heatmap} and \ref{fig:confusion}, where the vertical heatmap orders traces by anomaly score and reveals the narrow cluster of false positives expected at a 1\% FPR. Figures ~\ref{fig:embedding} and \ref{fig:training-curves} further confirms that discriminator embeddings place benign and tampered traces in distinct regions of feature space while the GAN training curves converge smoothly without instability.

\subsection{Timing And Throughput Considerations}
Capture (2{,}000 traces at 3{,}000 samples) took minutes on CW-Lite, and fewer traces would shorten time at some cost to accuracy. Training ran for tens of minutes on CPU for 400 epochs. Inference is lightweight (a forward pass through $D$) and fits inline screening once traces are captured. Compared to X-ray/CT or destructive analysis (hours to days per part), this is fast enough for a “tier-2” screen, and hardware acceleration or fewer traces can reduce per-unit time further.
We include training curves and per-trace visualizations (trace overlays and top anomalous traces) to show how optimization progressed and how anomalies manifest in the waveform domain.

\section{DISCUSSION}
\label{sec:discussion}

These results suggest a practical approach to supply chain screening that connects basic functional testing and expensive forensic analysis. The method's ability to operate without manufacturer cooperation or trusted reference hardware makes it applicable to commercial off-the-shelf components acquired through standard channels. By training only on benign examples, the approach avoids the need to anticipate specific attack implementations. The quantitative anomaly scores enable organizations to adjust sensitivity based on their available forensic capacity, creating a flexible screening tier that can adapt to different operational constraints and risk tolerances.

Both benign and tampered traces were collected in a single measurement session on one target microcontroller, which constrains the external validity of these results to other devices. Future validation across multiple devices, capture sessions, and fixture configurations is necessary to establish robustness for scalable deployment.
While the selective firmware builds provide per-payload insight, every experiment still relies on the SimpleSerial AES workload and a single trigger window. Expanding to additional microcontrollers and firmware as well as integrating complementary observables (timing, EM, or functional self-tests) are important directions for future work.

Microcontrollers routinely move through international manufacturing, packaging, and distribution channels before they arrive at board assembly lines and system integrators. That structure creates a persistent exposure wherein a component can be authentic at the point of manufacture and still become suspect later through substitution, remarking, refurbishment, or downstream modification. A scalable, non-destructive screen that can be run on incoming shipments and replacement parts gives buyers a way to introduce evidence-based checks into a supply chain that is otherwise governed largely by documentation and contractual assurances.

At the national level, this kind of screening capability supports the shift toward treating semiconductor integrity as a routine supply chain control rather than an exceptional forensic event. Current federal guidance emphasizes that supply chain risk management should be operationalized through repeatable controls and documented decisions across the lifecycle, but most organizations lack technical mechanisms that can be applied at volume. A practical screening tier strengthens enforcement of existing requirements by giving integrators and contractors a defensible basis to accept, quarantine, or escalate parts under real constraints, which is directly relevant to counterfeit avoidance expectations in defense procurement and to risk-based trusted- systems practice.

Internationally, the implication is not that screening replaces trusted supplier relationships, but that it makes them more resilient in a multi-tier market. Allies can align on shared screening practices at key chokepoints, such as incoming inspection at contract manufacturers, integration sites, and maintenance depots, so that integrity checks are not limited to a small set of high-end labs or applied only after an incident. This directly improves supply chain security by increasing coverage across large volumes of commodity components while preserving scarce forensic methods for the small subset of units that trigger concern. It also addresses a well documented policy problem, where counterfeit and suspect parts have entered high consequence supply chains before, and the damage is often discovered late, after integration, when remediation is most
expensive.

\section{CONCLUSION}
\label{sec:conclusion}

Current semiconductor supply chain security practices lack methods for detecting unauthorized hardware modifications between functional testing and expensive forensic analysis at scale. This paper demonstrates that  side-channel power measurements combined with one-class generative adversarial networks enable detection of tampered microcontrollers, achieving 93.2\% detection at 1\% false positive rate across multiple tampering scenarios. This approach provides a practical intermediate screening tier that directs expensive forensic resources toward anomalous devices, substantially increasing supply chain coverage without requiring comprehensive inspection of every component.


\acknowledgments
This work was supported by the U.S. Department of Defense (DoD) under Award No. HQ00342520002 as part of the Semiconductor Supply Chain and Cybersecurity Assessment program (CFDA 12.599).




\end{document}